\documentclass[a4,11pt]{article}
\usepackage[dvipdfmx]{pict2e}

\usepackage{lscape}
\usepackage{amsmath}
\usepackage{amssymb}
\usepackage{amsthm}
\usepackage{bm}
\usepackage{here}
\usepackage{mathtools}
\usepackage{newtxtext,newtxmath}
\usepackage{setspace}
\usepackage{url} 
\usepackage{ulem}
\usepackage{multirow}
\usepackage[dvipdfmx]{hyperref}
\hypersetup{
setpagesize=false,
 bookmarksnumbered=true,%
 bookmarksopen=true,%
 colorlinks=true,%
 linkcolor=blue,
 urlcolor=blue
}
\usepackage{framed}
\newtheorem{theorem}{Theorem}

\theoremstyle{remark}
\newtheorem{remark}{Remark}

\usepackage{amsmath,amssymb}

\usepackage{float}
\usepackage{amssymb}
\usepackage{latexsym}
\usepackage{amsmath}

\usepackage{amsmath,amssymb}
\DeclareMathOperator*{\argmax}{argmax}

\usepackage{graphicx}
\begin{document}
\fontsize{14pt}{20pt}\selectfont

\title{Small area estimation of dependent extreme value indices}
\author{{\sc{Koki Momoki}}$^{1}$\thanks{E-mail: momoki@sci.kagoshima-u.ac.jp} and {\sc{Takuma Yoshida}}$^{2}$\thanks{E-mail: yoshida@sci.kagoshima-u.ac.jp}\\\\
$^{1,2}${\it{Graduate School of Science and Engineering, Kagoshima University}}\\
{\it{1-21-40 Korimoto, Kagoshima, Kagoshima, 890-8580, Japan}}}

\date{}

\maketitle

\abstract{
\fontsize{12pt}{20pt}\selectfont
In extreme value analysis, tail behavior of a heavy-tailed data distribution is modeled by a Pareto-type distribution in which the so-called extreme value index (EVI) controls the tail behavior. For heavy-tailed data obtained from multiple population subgroups, or areas, this study efficiently predicts the EVIs of all areas using information among areas. For this purpose, we propose a mixed effects model, which is a useful approach in small area estimation. In this model, we represent differences among areas in the EVIs by latent variables called random effects. Using correlated random effects across areas, we incorporate the relations among areas into the model. The obtained model achieves simultaneous prediction of EVIs of all areas. Herein, we describe parameter estimation and random effect prediction in the model, and clarify theoretical properties of the estimator. Additionally, numerical experiments are presented to demonstrate the effectiveness of the proposed method. As an application of our model, we provide a risk assessment of heavy rainfall in Japan.
}

\bigskip

\noindent
{\it{
\fontsize{12pt}{20pt}\selectfont
}}

\bigskip

\noindent
{\it{
\fontsize{12pt}{20pt}\selectfont
}}

\fontsize{12pt}{20pt}\selectfont

\section{Introduction}\label{Section1}

Extreme value theory provides elegant statistical techniques for analyzing the probability of rare event occurrence in various fields. Examples include heavy rainfall, extreme temperatures, marked financial loss, and high medical costs. On this topic, numerous textbooks and review papers have been published, as presented by Gomes and Guillou (2015). Well-known approaches in extreme value analysis include the block maxima method and the peak-over-threshold (POT) method. The block maxima method uses the generalized extreme value distribution to model block maxima such as annual maxima. The POT method derives a tail model that fits high threshold exceedances to the generalized Pareto distribution (GPD). Mathematical details and connections between these two methods were discussed by de Haan and Ferreira (2006, Part I).

In the context of the POT method, the class of heavy-tailed distributions is described as the Pareto-type distribution rather than the GPD (Table 1 of Wang and Tsai 2009).
Numerous authors have studied the POT method for the Pareto-type distribution, including Gomes et al. (2008a, 2008b), Beirlant et al. (2009), and Girard et al. (2021).
The primary task of this method is to estimate the extreme value index (EVI), which is a parameter that dominates the right-tail behavior of the Pareto-type distribution.
Several estimators of the EVI have been proposed by Hill (1975), Dekker et al. (1989), and Gomes et al. (2008a).
Particularly the Hill estimator, proposed by Hill (1975), is a widely used estimator of the EVI.

This study examines the POT method for heavy-tailed data obtained from multiple population subgroups.
Each subgroup is referred to as an ``area'' (Rao and Molina 2015; Molina et al. 2022) for which data from the same area share common characteristics related to extreme events of interest.
The simplest strategy for predicting EVIs is to apply a classical method, such as the Hill estimator, directly to each area.
However, this area-wise analysis might be inefficient because the POT method only uses the high threshold exceedances, and the effective sample size of each area tends to be small in many applications.
To overcome this challenge, we aim to predict the EVIs of all areas efficiently using mutual information between areas.
In a similar context, Rohrbeck and Tawn (2019) proposed a clustering method for areas based on similarities in marginal effects.
However, this method leads to complete pooling of marginal distributions for some areas.
Such a method might be useful for easy interpretation, but it might fail to capture small differences between areas.
Unlike clustering, we do not assume homogeneity between areas or pool data from different areas.
As an alternative method of using inter-area information, we consider the use of a mixed effects model.

Mixed effect models have been studied within the framework of small area estimation.
Standard techniques of small area estimation are described by Jiang and Nguyen (2007), Wu (2009), and Jiang (2017).
This model is useful for producing reliable predictions across areas by borrowing information from covariates, especially in small sample settings (Sugasawa and Kubokawa 2020).
Then, differences in model parameters across areas are represented by latent variables called random effects.
These random effects are generally assumed to be independent among areas.
In the context of EVI regression, Momoki and Yoshida (2025) proposed the POT method using a mixed effects model.
However, their mixed effects model employed independent random effects with respect to areas.
Therefore, this study develops a method that uses correlated random effects and incorporates the relations among areas directly into the parameters.

The contributions and organization of this article are presented below.
Section \ref{Section2} introduces our POT method using the mixed effects model for the Pareto-type distribution.
Then, we reflect the relations between areas via correlated random effects.
For these correlated random effects, we construct the maximum likelihood method for estimating parameters and the conditional mode method for predicting random effects.
We can predict the EVIs by combining these two methods.
Section \ref{Section3} investigates the mathematical properties of the proposed estimator.
Under the use of independent random effects, Nie (2007) and Jiang et al. (2022) have established the asymptotic normality of the maximum likelihood estimator for generalized mixed effects models, whereas Momoki and Yoshida (2025) extended their work to the POT method.
The asymptotic theory in Section \ref{Section3} covers correlated random effects. It can be regarded as a novel result even within the context of generalized mixed effects models as well as extreme value theory.
The obtained asymptotic normality indicates that the variance of the proposed estimator decreases as the number of areas increases.
This property supports our motivation to incorporate more information among areas.
For the proposed method, Section \ref{Section4} revealed its characteristics of numerical performance through a simulation study.
We verified that our model improves area-wise estimates markedly by appropriately setting correlations of random effects.
Therefore, Section \ref{Section5} presents several examples of incorporating relations among areas as correlations of random effects.
Section \ref{Section6} applies the proposed model to a real dataset of extreme precipitation in Japan.
The dataset includes records from 1138 stations (i.e., areas).
For this large dataset, the proposed method produced more reasonable results than the area-wise estimation method did.
Section \ref{Section7} summarizes the main points of this paper. The proof of the theorem in Section \ref{Section3} is given in the Appendix.

\section{Mixed Effects Modeling of Tail Probability}\label{Section2}

We consider data observed from multiple areas.
Letting $J\in\mathbb{N}$ be the number of areas, we identify the areas by applying the labels $\mathcal{J}\coloneqq\{1, 2, \ldots, J\}$.
Let
\begin{equation}
\left\{Y_{ij}\in\mathbb{R},\ i=1, 2, \ldots, n_j,\ j\in\mathcal{J}\right\}\label{Eq2.0.1}
\end{equation}
be an independent random sample from the areas $\mathcal{J}$, where $n_j$ is the sample size for the area $j\in\mathcal{J}$, and $Y_{ij}$ is an $i$-th observation in the area $j\in\mathcal{J}$.
For (\ref{Eq2.0.1}), we assume that data from the same area have the same area-specific underlying distribution.
As described in Section \ref{Section1}, we aim to predict the right tails of the underlying distributions for all areas $\mathcal{J}$ using information about the relations between the areas.
To this end, we develop the POT method using the mixed effects model.
Section \ref{Section2.1} presents the simplest approach of the POT method to (\ref{Eq2.0.1}).
Subsequently, Sections \ref{Section2.2} and \ref{Section2.3} explain our proposed POT method.

\subsection{Classical approach of the peak-over-threshold method}\label{Section2.1}

Let $F_j(y)\coloneqq P(Y_{ij}\leq y)$ be the distribution function of the data $\{Y_{ij}\}_{i=1}^{n_j}$ obtained from the area $j\in\mathcal{J}$.
The typical interest in extreme value analysis is to evaluate the right tails of the underlying distributions $F_j(y),\ j\in\mathcal{J}$ accurately, where these targets can be expressed as
\begin{equation}
\begin{split}\label{Eq2.1.1}
F_j(y\mid \omega_j)&\coloneqq P(Y_{ij}\leq y\mid Y_{ij}>\omega_j)\\
&=\frac{F_j(y)-F_j(\omega_j)}{1-F_j(\omega_j)},\quad y>\omega_j,\ j\in\mathcal{J}
\end{split}
\end{equation}
for some high thresholds $\{\omega_j\}_{j\in\mathcal{J}}$.
For this purpose, the use of the following approximation of (\ref{Eq2.1.1}) is standard for extreme value analysis (Smith 1987; Drees et al. 2004):
\begin{equation}
F_j(y\mid \omega_j)\approx 1-\left(1+\gamma_j\frac{y-\omega_j}{\psi_j}\right)_+^{-1/\gamma_j},\quad j\in\mathcal{J}.\label{Eq2.1.2}
\end{equation}
Therein, the right-hand side of (\ref{Eq2.1.2}) represents the distribution function of the GPD with a location parameter $\omega_j\in\mathbb{R}$, scale parameter $\psi_j>0$ and shape parameter $\gamma_j\in\mathbb{R}$, and $x_+\coloneqq\max\{0, x\}$ for $x\in\mathbb{R}$.
When $\gamma_j=0$, the right-hand side of (\ref{Eq2.1.2}) means $1-\exp[-(y-\omega_j)/\psi_j]$, which is the exponential distribution.
According to (\ref{Eq2.1.2}), the targets $F_j(y\mid \omega_j),\ j\in\mathcal{J}$ are predictable by fitting the high threshold exceedances $\{Y_{ij}: Y_{ij}>\omega_j,\ i=1, 2, \ldots, n_j,\ j\in\mathcal{J}\}$ to the GPD and by estimating the unknown parameters of the GPD.
This approach, called the POT method, is commonly justified by extreme value theory under the high thresholds $\{\omega_j\}_{j\in\mathcal{J}}$ (Chapters 1 and 3 of de Haan and Ferreira 2006).

This study examines a special type of the POT method presented above.
Specifically, we assume that the underlying distributions $F_j(y),\ j\in\mathcal{J}$ belong to a class of distributions called Pareto-type distributions (Wang and Tsai 2009).
Therefore, $F_j(y),\ j\in\mathcal{J}$ can be written as
\begin{equation}
F_j(y)=1-y^{-1/\gamma_j}\mathcal{L}_j(y),\quad y>0,\ j\in\mathcal{J},\label{Eq2.1.3}
\end{equation}
where $\gamma_j>0$ is an unknown parameter called the EVI, and $\mathcal{L}_j(y)$ is a slowly varying function, meaning that for any $s>0$, $\mathcal{L}_j(ys)/\mathcal{L}_j(y)\to 1$ as $y\to\infty$.
As shown in Table 1 of Wang and Tsai (2009), the Pareto-type distribution (\ref{Eq2.1.3}) covers many heavy-tailed distributions.
Under the form (\ref{Eq2.1.3}), the right-tail distributions (\ref{Eq2.1.1}) are approximated (Wang and Tsai 2009) as
\begin{equation}
F_j(y\mid \omega_j)\approx 1-\left(\frac{y}{\omega_j}\right)^{-1/\gamma_j},\quad y>\omega_j,\ j\in\mathcal{J}.\label{Eq2.1.4}
\end{equation}
Approximation (\ref{Eq2.1.4}) is well-known to be closely related to (\ref{Eq2.1.2}) with $\gamma_j>0,\ j\in\mathcal{J}$ (Theorem 1.2.1 of de Haan and Ferreira 2006).
In the model (\ref{Eq2.1.4}), the targets $F_j(y\mid \omega_j),\ j\in\mathcal{J}$ can be detected if the parameters $\{\gamma_j\}_{j\in\mathcal{J}}$ can be estimated.

A widely used estimator of $\gamma_j>0$ is the Hill estimator (Hill 1975):
\begin{equation}
\hat{\gamma}_j^{\rm{H}}\coloneqq\frac{\sum_{i=1}^{n_j}\log (Y_{ij}/\omega_j)I(Y_{ij}>\omega_j)}{\sum_{i=1}^{n_j}I(Y_{ij}>\omega_j)},\label{Eq2.1.5}
\end{equation}
where $I(\cdot)$ is an indicator function defined as
\begin{equation*}
I(Y_{ij}>\omega_j)\coloneqq\begin{cases}
0, & Y_{ij}\leq \omega_j,\\
1, & Y_{ij}>\omega_j.\label{Eq2.1.6}
\end{cases}
\end{equation*}
The classical estimators $\{\hat{\gamma}_j^{\rm{H}}\}_{j\in\mathcal{J}}$, which are denoted by area-wise estimators, use no mutual information between areas.
It is noteworthy that each $\hat{\gamma}_j^{\rm{H}}$ is constructed by only the threshold exceedances $\{Y_{ij}: Y_{ij}>\omega_j,\ i=1, 2, \ldots, n_j\}$. Therefore, its effective sample size is usually small under the setting of the high threshold $\omega_j$.
Consequently, area-wise estimation in extreme value analysis might not be effective.
To obtain more efficient estimates, we incorporate the relations between the areas $\mathcal{J}$ into (\ref{Eq2.1.3}) and reflect the information from all areas $\mathcal{J}$ into a single model.
In Section \ref{Section2.2}, such a sophisticated model for (\ref{Eq2.1.3}) is developed using the mixed effects model, which is widely used in small area estimation (Jiang and Nguyen 2007; Wu 2009).

\subsection{Utilization of mixed effects models}\label{Section2.2}

The mixed effects model captures differences between the areas in the distributions of the data through latent variables called random effects, rather than parameters.
Let $V_1, V_2, \ldots, V_J$ represent random effects that are marginally distributed as normal $V_j\sim N(0, \sigma^2),\ j\in\mathcal{J}$ with mean zero and unknown variance $\sigma^2>0$.
The use of the Gaussian distribution for the random effects is standard in the mixed effects model (Jiang and Nguyen 2007; Wu 2009).

We assume that the EVIs $\{\gamma_j\}_{j\in\mathcal{J}}$ are represented by
\begin{equation}
\gamma_j=\exp(\mu+V_j)>0,\quad j\in\mathcal{J},\label{Eq2.2.1}
\end{equation}
where $\mu\in\mathbb{R}$ is an unknown common parameter across all areas $\mathcal{J}$ (Remark \ref{Remark1}).
Momoki and Yoshida (2025) also used the exponential link function to ensure positivity on the right-hand side of (\ref{Eq2.2.1}).
Unlike the classical approach described in Section \ref{Section2.1}, we model the relations between the areas $\mathcal{J}$ in the EVIs $\{\gamma_j\}_{j\in\mathcal{J}}$.
To accomplish this improvement, our random effects $\{V_j\}_{j\in\mathcal{J}}$ can be correlated as
\begin{equation*}
D_{j_1j_2}\coloneqq Cor[V_{j_1}, V_{j_2}]\in[-1, 1],\quad j_1, j_2\in\mathcal{J}\label{Eq2.2.2}
\end{equation*}
and thus satisfy
\begin{equation}
{\bm{V}}\coloneqq(V_1, V_2, \ldots, V_J)^\top\sim N_J({\bm{0}}, \sigma^2{\bm{D}})\label{Eq2.2.3}
\end{equation}
for a known positive definite correlation matrix ${\bm{D}}\coloneqq[D_{j_1j_2}]_{j_1, j_2\in\mathcal{J}}$ (Remark \ref{Remark2}). 
Note that the diagonal entries of ${\bm{D}}$ are set to 1.
Roughly speaking, if two areas $j_1\in\mathcal{J}$ and $j_2\in\mathcal{J}$ have a strong positive correlation $D_{j_1j_2}$ of the random effects, then their EVIs $\gamma_{j_1}=\exp(\mu+V_{j_1})$ and $\gamma_{j_2}=\exp(\mu+V_{j_2})$ are likely to have similar values.
In this sense, the EVIs (\ref{Eq2.2.1}) with the correlated random effects (\ref{Eq2.2.3}) can reflect the relations among areas $\mathcal{J}$.

The marginal distribution of the area $j\in\mathcal{J}$ is expressed as $F_j(y\mid v_j)\coloneqq P(Y_{ij}<y \mid V_j=v_j)$.
According to (\ref{Eq2.1.3}), $F_j(y\mid v_j)$ can be modeled as
\begin{equation}
F_j(y\mid v_j)=1-y^{-1/\exp(\mu+v_j)}\mathcal{L}_j(y; v_j),\quad y>0,\ j\in\mathcal{J},\label{Eq2.2.4}
\end{equation}
where $\exp(\mu+v_j)$ represents the EVI as explained in (\ref{Eq2.2.1}), and $\mathcal{L}_j(y; v_j)$ is a slowly varying function with respect to $y$.
As described in (\ref{Eq2.2.3}), we use the correlated random effects $\{V_j\}_{j\in\mathcal{J}}$ for this model, which are useful for representing the relations among the areas $\mathcal{J}$ in (\ref{Eq2.2.4}) (Remark \ref{Remark3}).

\bigskip

\begin{remark}\label{Remark1}
The proposed model (\ref{Eq2.2.1}) for the EVI can be assumed to follow a log-normal distribution.
The log-normal distribution is well known to be asymmetric, with a mode of $\exp(\mu-\sigma^2)$, a median of $\exp(\mu)$, and a mean of $\exp(\mu+\sigma^2/2)$.
Therefore, the model (\ref{Eq2.2.1}) tends to support larger values of the EVI.
Because a larger EVI indicates a heavier right tail of the underlying distribution, this structure for the EVI suggests a conservative approach to risk prediction.
Consequently, our model (\ref{Eq2.2.1}) is reasonable for avoiding underestimation of extreme events.
\end{remark}

\bigskip

\begin{remark}\label{Remark2}
The correlation matrix ${\bm{D}}$ of the random effects $\{V_j\}_{j\in\mathcal{J}}$ is determined by the $J(J-1)/2$ components.
Consequently, if $J$ is large and all components of ${\bm{D}}$ are completely unknown, then the number of parameters to be estimated increases drastically.
For the model (\ref{Eq2.2.4}), such numerous unknown parameters of ${\bm{D}}$ might engender uncertain estimates of the other parameters $\mu$ and $\sigma^2$.
To avoid this difficulty, the construction of ${\bm{D}}$ is separate from the estimation procedure of $(\mu,\sigma^2)$.
Examples of constructing ${\bm{D}}$ are discussed in Section \ref{Section5}.
\end{remark}

\bigskip

\begin{remark}\label{Remark3}
In general, area-wise estimation leads to poor performance for areas with small sample sizes (Diallo and Rao 2018).
The goal of small area estimation is to improve area-wise estimates for areas with small sample sizes. The mixed effects model is useful for this purpose (Sugasawa and Kubokawa 2020).
Typically, the random effects of the mixed effects model are assumed to be independent across areas (Nie 2007, Jiang et al. 2022).
This assumption implies that the correlation matrix ${\bm{D}}$ of the random effects is fixed as the identity matrix ${\bm{I}}_J$.
In the context of EVI regression, Momoki and Yoshida (2025) demonstrated the effectiveness of independent random effects.
Unlike their work, this study is novel in that we can employ the non-identity matrix ${\bm{D}}\neq{\bm{I}}_J$.
By this modification, our mixed effects model (\ref{Eq2.2.4}) can incorporate the similarities among areas $\mathcal{J}$ in the EVIs $\{\gamma_j\}_{j\in\mathcal{J}}$.
In this sense, the proposed model (\ref{Eq2.2.4}) achieves ``borrowing of strength'' between the areas $\mathcal{J}$ (Dempster et al. 1981).
Sections \ref{Section5} and \ref{Section6} present some examples of constructing ${\bm{D}}\neq{\bm{I}}_J$ and show verification of the performance of the proposed model for these examples.
\end{remark}

\subsection{Estimation and prediction methods}\label{Section2.3}

The unknown parameters $\mu$ and $\sigma^2$ are estimated using the maximum likelihood method.
Furthermore, to capture differences between the areas $\mathcal{J}$ in the EVIs $\{\gamma_j\}_{j\in\mathcal{J}}$, we must also predict the random effects $\{V_j\}_{j\in\mathcal{J}}$.
Random effects $\{V_j\}_{j\in\mathcal{J}}$ are predicted using the conditional mode method.
These estimation and prediction methods are standard in generalized mixed effects models (Section 3.6.2 of Jiang and Nguyen 2007; Chapter 11 of Wu 2009).
Detailed definitions of our estimator and predictor are given in the following Sections \ref{Section2.3.1} and \ref{Section2.3.2}.
We can implement these proposed methods using the Template Model Builder ({\textsf{TMB}}) package in the {\textsf{R}} environment (Kristensen et al. 2016).

\subsubsection{Approximate maximum likelihood estimation}\label{Section2.3.1}

Assuming that under conditioning on ${\bm{V}}=(V_1, V_2, \ldots, V_J)^\top$, then the data $\{Y_{ij},\ i=1, 2, \ldots, n_j,\ j\in\mathcal{J}\}$ are independent both within the same area and between the areas $\mathcal{J}$ (Jiang et al. 2022).
According to the standard definition of the likelihood function for mixed effects models (Chapter 2 of Wu 2009), the likelihood for the threshold exceedances $\{Y_{ij}: Y_{ij}>\omega_j,\ i=1, 2, \ldots, n_j,\ j\in\mathcal{J}\}$ is definable by
\begin{equation*}
E_{\bm{V}}\left[\prod_{j\in\mathcal{J}}\prod_{i=1, 2, \ldots, n_j: Y_{ij}>\omega_j}\frac{\partial}{\partial y}F_j(y\mid \omega_j, V_j)\mid_{y=Y_{ij}}\right],\label{Eq2.3.1.1}
\end{equation*}
which involves the expectation over the random effects distribution because the random effects $\{V_j\}_{j\in\mathcal{J}}$ are latent variables, where
\begin{equation*}
\begin{split}\label{Eq2.3.1.2}
F_j(y\mid \omega_j, v_j)&\coloneqq P(Y_{ij}\leq y\mid Y_{ij}>\omega_j, V_j=v_j)\\
&=\frac{F_j(y\mid v_j)-F_j(\omega_j\mid v_j)}{1-F_j(\omega_j\mid v_j)},\quad y>\omega_j,\ j\in\mathcal{J}.
\end{split}
\end{equation*}
Similarly to (\ref{Eq2.1.4}), under the model (\ref{Eq2.2.4}) and assumption (A1) of Section \ref{Section3}, the probability density function $(\partial/\partial y)F_j(y\mid \omega_j, v_j)$ can be approximated as
\begin{equation}
\begin{split}\label{Eq2.3.1.3}
\frac{\partial}{\partial y}F_j(y\mid \omega_j, v_j)&\approx{\rm{pareto}}_{\omega_j}(y; \exp(\mu+v_j))\\
&\coloneqq\frac{1}{\omega_j\exp(\mu+v_j)}\left(\frac{y}{\omega_j}\right)^{-1/\exp(\mu+v_j)-1},\quad y>\omega_j,\ j\in\mathcal{J},
\end{split}
\end{equation}
where ${\rm{pareto}}_{\omega_j}(y; \exp(\mu+v_j))$ represents the density of the Pareto distribution of $Y_{ij}/\omega_j$ with parameter $\exp(\mu+v_j)$.
The same approximation as (\ref{Eq2.3.1.3}) was also provided by Wang and Tsai (2009, Section 2.2).
Therefore, the likelihood function for estimating the unknown parameters $\mu$ and $\sigma^2$ is obtained approximately as
\begin{equation}
\begin{split}\label{Eq2.3.1.4}
L(\mu, \sigma^2)&\coloneqq E_{\bm{V}}\left[\prod_{j\in\mathcal{J}}\prod_{i=1, 2, \ldots, n_j: Y_{ij}>\omega_j}{\rm{pareto}}_{\omega_j}(Y_{ij}; \exp(\mu+v_j))\right]\\
&=\int_{\mathbb{R}^J}\phi_J({\bm{v}}; {\bm{0}}, \sigma^2{\bm{D}})\prod_{j\in\mathcal{J}}\prod_{i=1, 2, \ldots, n_j: Y_{ij}>\omega_j}{\rm{pareto}}_{\omega_j}(Y_{ij}; \exp(\mu+v_j))d{\bm{v}},
\end{split}
\end{equation}
where ${\bm{v}}\coloneqq(v_1, v_2, \ldots, v_J)^\top\in\mathbb{R}^J$, and $\phi_J({\bm{v}}, {\bm{0}}, \sigma^2{\bm{D}})$ is the joint probability density function of $N_J({\bm{0}}, \sigma^2{\bm{D}})$.
From the likelihood function $L(\mu, \sigma^2)$ presented above, we can derive the maximum likelihood estimators of $\mu$ and $\sigma^2$.
We denote these estimators as $\hat{\mu}\in\mathbb{R}$ and $\hat{\sigma}^2>0$.

\subsubsection{Conditional mode method for predicting random effects}\label{Section2.3.2}

According to Wu (2009), the conditional joint density function of ${\bm{V}}=(V_1, V_2, \ldots, V_J)^\top$ given $\{Y_{ij}: Y_{ij}>\omega_j,\ i=1, 2, \ldots, n_j,\ j\in\mathcal{J}\}$ is proportional to
\begin{equation}
\phi_J({\bm{v}}; {\bm{0}}, \sigma^2{\bm{D}})\prod_{j\in\mathcal{J}}\prod_{i=1, 2, \ldots, n_j: Y_{ij}>\omega_j}\frac{\partial}{\partial y}F_j(y\mid \omega_j, v_j)\mid_{y=Y_{ij}},\label{Eq2.3.2.1}
\end{equation}
which is the function of ${\bm{v}}=(v_1, v_2, \ldots, v_J)^\top$.
Similarly to Section \ref{Section2.3.1}, we again approximate $(\partial/\partial y)F_j(y\mid \omega_j, v_j)$ by the Pareto density ${\rm{pareto}}_{\omega_j}(y; \exp(\mu+v_j))$.
Then, the empirical version of (\ref{Eq2.3.2.1}) is given as
\begin{equation}
\phi_J({\bm{v}}; {\bm{0}}, \hat{\sigma}^2{\bm{D}})\prod_{j\in\mathcal{J}}\prod_{i=1, 2, \ldots, n_j: Y_{ij}>\omega_j}{\rm{pareto}}_{\omega_j}(Y_{ij}; \exp(\hat{\mu}+v_j)),\label{Eq2.3.2.2}
\end{equation}
where $\hat{\mu}$ and $\hat{\sigma}^2$ respectively represent the maximum likelihood estimators of $\mu$ and $\sigma^2$.
The predictor of ${\bm{V}}=(V_1, V_2, \ldots, V_J)^\top$ is defined by the mode of (\ref{Eq2.3.2.2}) as
\begin{equation*}
\begin{split}\label{Eq2.3.2.3}
\tilde{\bm{v}}&=(\tilde{v}_1, \tilde{v}_2, \ldots, \tilde{v}_J)^\top\\
&\coloneqq\underset{{\bm{v}}\in\mathbb{R}^J}{\rm{\argmax}}\ \phi_J({\bm{v}}; {\bm{0}}, \hat{\sigma}^2{\bm{D}})\prod_{j\in\mathcal{J}}\prod_{i=1, 2, \ldots, n_j: Y_{ij}>\omega_j}{\rm{pareto}}_{\omega_j}(Y_{ij}; \exp(\hat{\mu}+v_j))
\end{split}
\end{equation*}
(Remark \ref{Remark4}).
Particularly, the EVIs $\{\gamma_j\}_{j\in\mathcal{J}}$ are predicted as
\begin{equation}
\tilde{\gamma}_j\coloneqq\exp\left(\hat{\mu}+\tilde{v}_j\right),\quad j\in\mathcal{J}.\label{Eq2.3.2.4}
\end{equation}

\bigskip

\begin{remark}\label{Remark4}
The log-transformation of (\ref{Eq2.3.2.2}) is given as
\begin{equation}
\log\phi_J({\bm{v}}; {\bm{0}}, \hat{\sigma}^2{\bm{D}})+\sum_{j\in\mathcal{J}}\sum_{i=1, 2, \ldots, n_j: Y_{ij}>\omega_j}\log{\rm{pareto}}_{\omega_j}(Y_{ij}; \exp(\hat{\mu}+v_j)).\label{Eq2.3.2.5}
\end{equation}
For simplicity, we examine the structure of (\ref{Eq2.3.2.5}) with $J=2$, ${\bm{V}}=(V_1, V_2)^\top$, ${\bm{v}}=(v_1, v_2)^\top$, $\tilde{\bm{v}}=(\tilde{v}_1, \tilde{v}_2)^\top$, and $Cor[V_1, V_2]=\rho$.
The first term of (\ref{Eq2.3.2.5}), i.e., $\log\phi_J({\bm{v}}; {\bm{0}}, \hat{\sigma}^2{\bm{D}})$ depends on ${\bm{v}}$ through
\begin{equation}
-\frac{{\bm{v}}^\top{\bm{D}}^{-1}{\bm{v}}}{2\hat{\sigma}^2}=-\frac{\left(v_1-v_2\right)^2}{2\hat{\sigma}^2\left(1-\rho^2\right)} -\frac{v_1 v_2}{\hat{\sigma}^2\left(1+\rho\right)}.\label{Eq2.3.2.6}
\end{equation}
Then, we have $\log\phi_J({\bm{v}}; {\bm{0}}, \hat{\sigma}^2{\bm{D}})\to-\infty$ as $\rho\to1$, whereas the second term of (\ref{Eq2.3.2.5}) does not change directly with $\rho$. 
Consequently, for the maximization of (\ref{Eq2.3.2.5}), when $\rho\approx 1$, the effect of $\log\phi_J({\bm{v}}; {\bm{0}}, \hat{\sigma}^2{\bm{D}})$ is far greater than that of the second term of (\ref{Eq2.3.2.5}). Then, $v_1$ and $v_2$ should be mutually close to increase $\log\phi_J({\bm{v}}; {\bm{0}}, \hat{\sigma}^2{\bm{D}})$.
Therefore, the use of $\rho=Cor[V_1, V_2]\approx1$ suggests similar values of $\tilde{v}_1$ and $\tilde{v}_2$. 
If $\rho=0$, then $\tilde{v_1}$ and $\tilde{v_2}$ are obtained independently, although their magnitudes are restricted by $\hat{\sigma}^2$.
Consequently, because of the correlation structure of ${\bm{D}}$, $(v_1,\ldots,v_J)$ are optimized in a mutually dependent manner.
\end{remark}

\section{Asymptotic Theory}\label{Section3}
\subsection{Conditions}\label{Section3.1}

We develop an asymptotic theory for the proposed model (\ref{Eq2.2.4}) under the conditions $n_j\to\infty,\ j\in\mathcal{J}$ and $J\to\infty$.
Thus, the number of areas, $J$, as well as all sample sizes $\{n_j\}_{j\in\mathcal{J}}$ are assumed to be sufficiently large.
The same conditions are discussed by Jiang et al. (2022).

For each area $j\in\mathcal{J}$, $\omega_j$ is assumed to be a sequence of $n_j$ such that $\omega_j\to\infty$ as $n_j\to\infty$, which is standard in extreme value theory (Smith 1987).
We denote $k_j\coloneqq \sum_{i=1}^{n_j}I(Y_{ij}>\omega_j),\ j\in\mathcal{J}$.
Then, $k_j$ represents the effective sample size for the area $j\in\mathcal{J}$.
The average of the area-wise effective sample sizes $\{k_j\}_{j\in\mathcal{J}}$ is denoted by $k\coloneqq J^{-1}\sum_{j\in\mathcal{J}}k_j$.
Our asymptotic theory requires that the following conditions (A1)--(A5) hold uniformly for all $v_j\in\mathbb{R}$ and all $j\in\mathcal{J}$:
\begin{itemize}
\item[(A1)] The slowly varying function $\mathcal{L}_j(y; v_j)$ in (\ref{Eq2.2.4}) belongs to the Hall class (Hall 1982).
In other words, it is represented as
\begin{equation*}
\mathcal{L}_j(y; v_j)=a_j(v_j) + b_j(v_j)y^{-\beta_j(v_j)}+o\left(y^{-\beta_j(v_j)}\right),\label{Eq3.1.1}
\end{equation*}
where $a_j(\cdot)>0$, $b_j(\cdot)$ and $\beta_j(\cdot)>0$ are continuous and bounded functions.
\item[(A2)] Under conditioning on $V_j=v_j$, $k_j^{-1}\xrightarrow{P}0$ as $n_j\to\infty,\ j\in\mathcal{J}$ and $J\to\infty$, where ``$\xrightarrow{P}$'' stands for convergence in probability.
\item[(A3)] There exists a function $\theta_j: \mathbb{R}^J\to\mathbb{R}$ and some constants $ 0<\tau_1<\tau_2<\infty$ such that under conditioning on ${\bm{V}}={\bm{v}}$, $k_j/k\xrightarrow{P}\theta_j({\bm{v}})\in(\tau_1, \tau_2)$ as $n_j\to\infty,\ j\in\mathcal{J}$ and $J\to\infty$.
\item[(A4)] $k/J\xrightarrow{P} 0$ as $n_j\to\infty,\ j\in\mathcal{J}$ and $J\to\infty$.
\item[(A5)] There exist some constants $h_\mu\in\mathbb{R}$ and $h_{\sigma^2}\in\mathbb{R}$ such that
\begin{equation*}
\left({\bm{1}}_J^\top{\bm{D}}^{-1}{\bm{1}}_J\right)^{-1/2}{\bm{1}}_J^\top{\bm{D}}^{-1}E\left[\left[\varepsilon_j(V_j)\right]_{j\in\mathcal{J}}\right]\to h_{\mu}
\end{equation*}
and
\begin{equation*}
2J^{-1/2}E\left[{\bm{V}}^\top{\bm{D}}^{-1}\left[\varepsilon_j(V_j)\right]_{j\in\mathcal{J}}\right]\to h_{\sigma^2}
\end{equation*}
as $n_j\to\infty,\ j\in\mathcal{J}$ and $J\to\infty$, where ${\bm{1}}_J\coloneqq(1, 1, \ldots, 1)^\top\in\mathbb{R}^J$, and
\begin{equation*}
\varepsilon_j(V_j)\coloneqq\frac{b_j(V_j)\exp(\mu+V_j)\beta_j(V_j)}{1+\exp(\mu+V_j)\beta_j(V_j)}\omega_j^{-1/\exp(\mu+V_j)-\beta_j(V_j)},\quad j\in\mathcal{J}.
\end{equation*}
\end{itemize}
First, (A1) regularizes the convergence of $F_j(y\mid \omega_j, v_j)$ to the Pareto distribution (Section 2.3 of de Haan and Ferreira 2006).
This condition is called the second-order condition.
Second, (A2)--(A4) are assumptions about the effective sample sizes $\{k_j\}_{j\in\mathcal{J}}$. They control the divergence rates of the thresholds $\{\omega_j\}_{j\in\mathcal{J}}$.
A detailed explanation of each of (A2)--(A4) is presented in Section 3.1 of Momoki and Yoshida (2025).
(A5) shows the rates at which the biases of the estimators are removed asymptotically, which is necessary in the asymptotic theory for technical reasons.

\subsection{Asymptotic properties}\label{Section3.2}

For the proposed estimators $\hat{\mu}$ and $\hat{\sigma}^2$, we obtain the following Theorem \ref{Theorem1}.

\begin{theorem}\label{Theorem1}
Suppose that (A1)--(A5) hold.
Then, as $n_j\to\infty,\ j\in\mathcal{J}$ and $J\to\infty$,
\begin{equation*}
\begin{bmatrix}
\sqrt{{\bm{1}}_J^\top{\bm{D}}^{-1}{\bm{1}}_J}\left(\hat{\mu}-\mu\right)\\
\sqrt{J}\left(\hat{\sigma}^2-\sigma^2\right)
\end{bmatrix}
-\begin{bmatrix}
h_\mu\\
h_{\sigma^2}
\end{bmatrix}
\xrightarrow{D}N\left(
\begin{bmatrix}
0\\
0
\end{bmatrix},\ 
\begin{bmatrix}
\sigma^2 & 0\\
0 & 2\sigma^4
\end{bmatrix}\right).\label{Eq3.2.1}
\end{equation*}
\end{theorem}

\bigskip

Some remarks about Theorem \ref{Theorem1} above are presented below.
\begin{itemize}
\item[(R1)] Jiang et al. (2022) studied the asymptotic normality of the maximum likelihood estimator for the generalized mixed effects model.
Momoki and Yoshida (2025) extended the result of Jiang et al. (2022) to the Pareto-type distribution in the context of extreme value theory.
Although Jiang et al. (2022) and Momoki and Yoshida (2025) assumed that the random effects are independent across areas, i.e., $\bm{D}={\bm{I}}_J$, the random effects $\{V_j\}_{j\in\mathcal{J}}$ in this study can have correlation between the areas as (\ref{Eq2.2.3}).
Therefore, our Theorem \ref{Theorem1} is novel in that it establishes a more general theory for random effects that may not be independent.
\item[(R2)] From Theorem \ref{Theorem1}, $\hat{\mu}$ is $\sqrt{{\bm{1}}_J^\top{\bm{D}}^{-1}{\bm{1}}_J}$-consistent.
If ${\bm{D}}={\bm{I}}_J$, then its convergence rate is $O(\sqrt{J})$, which corresponds to the slowest case.  
However, if all elements of ${\bm{D}}$ are non-zero and strongly correlated, then the rate becomes fastest, i.e., $O(J)$.  
Therefore, incorporating more areas into the model is expected to improve the estimator considerably. This property might be especially notable when ${\bm{D}}\neq{\bm{I}}_J$.
\item[(R3)] In Theorem \ref{Theorem1}, $h_{\mu}$ and $h_{\sigma^2}$ arise from the operation of peak-over-threshold using the approximation (\ref{Eq2.3.1.3}).
These are quantified by the second-order condition (A1).
If $h_{\mu}$ and $h_{\sigma^2}$ are large, then the estimators $\hat{\mu}$ and $\hat{\sigma}^2$ might remain biased as $n_j\to\infty,\ j\in\mathcal{J}$ and $J\to\infty$.
If we wish to correct the biases of the estimators, then we especially need to estimate the second-order parameters (functions) $\{\beta_j(v_j)\}_{j\in\mathcal{J}}$ appeared in (A1).
Although several studies (Gomes et al. 2002; de Wet et al. 2012) have explored estimation of the second-order parameter, methods for estimating it in the context of the mixed effects model have not yet been developed.
However, the standard deviations of $\hat{\mu}$ and $\hat{\sigma}^2$ remain stable when $J$ is large.
In this sense, it is important that the asymptotic rates of $\hat{\mu}$ and $\hat{\sigma}^2$ be dominated by $J$ instead of the effective sample sizes $\{k_j\}_{j\in\mathcal{J}}$.
Therefore, as one benefit of using the mixed effects model, the biases in the proposed estimators can be relaxed by setting higher thresholds $\{\omega_j\}_{j\in\mathcal{J}}$.
\end{itemize}

\section{Simulation Study}\label{Section4}

Through a simulation study, we confirmed the benefits of the proposed mixed effects model (\ref{Eq2.2.4}).
We present the results of a comparison between our proposed method (\ref{Eq2.3.2.4}) and the area-wise estimation method using the Hill estimator (\ref{Eq2.1.5}).

\subsection{Data generation procedure}\label{Section4.1}

Throughout this simulation study, we set the number of areas as $J=1000$.
We fixed the EVIs $\{\gamma_j\}_{j\in\mathcal{J}}$ as
\begin{equation}
\gamma_j=2\left(\frac{j-1}{J-1}-\frac{1}{2}\right)^2+\frac{1}{5},\quad j\in\mathcal{J}.\label{Eq4.1.1}
\end{equation}
Using these EVIs, the dataset $\{Y_{ij},\ i=1, 2, \ldots, n_j,\ j\in\mathcal{J}\}$ was simulated as an i.i.d. random sample from the Pareto distributions as
\begin{equation*}
P(Y_{ij}<y)=1-y^{-1/\gamma_j},\quad j\in\mathcal{J}.\label{Eq4.1.2}
\end{equation*}
To evaluate the performance of our proposed method and the Hill estimator, we generated $M=100$ replicates of the dataset above and predicted the EVIs $\{\gamma_j\}_{j\in\mathcal{J}}$ for each replicate. 
For simplicity, both methods employed the fixed thresholds $\omega_j=1,\ j\in\mathcal{J}$, implying $n_j=k_j,\ j\in\mathcal{J}$.
Under this simple situation, we considered the two cases $n\coloneqq n_1=n_2=\cdots=n_J=50$ and $n\coloneqq n_1=n_2=\cdots=n_J=200$.

\subsection{Results}\label{Section4.2}

To use our proposed method (\ref{Eq2.3.2.4}), we must first determine ${\bm{D}}=[D_{j_1j_2}]_{j_1, j_2\in\mathcal{J}}$ (Section \ref{Section5}).
The true EVIs (\ref{Eq4.1.1}) are similar in areas with similar area labels.
We assumed this information was known, and then employed
\begin{equation}
D_{j_1, j_2}=\exp\left(-\frac{\left\lvert j_1-j_2\right\rvert}{500}\right),\quad j_1, j_2\in\mathcal{J}\label{Eq4.2.1}
\end{equation}
and
\begin{equation}
D_{j_1, j_2}=\exp\left(-\frac{\left\lvert j_1-j_2\right\rvert}{1000}\right),\quad j_1, j_2\in\mathcal{J}.\label{Eq4.2.2}
\end{equation}
Even if we have no information about ${\bm{D}}$, our method is useful for ${\bm{D}}={\bm{I}}_J$.
Therefore, we also considered ${\bm{D}}={\bm{I}}_J$.
The three setups (\ref{Eq4.2.1}), (\ref{Eq4.2.2}), and ${\bm{D}}={\bm{I}}_J$ are designated respectively as ``Case 1'', ``Case 2'', and ``Case 3''.

\begin{table}[t]
\centering
\caption{Prediction results obtained using our method (\ref{Eq2.3.2.4}) and using the Hill estimator (\ref{Eq2.1.5}).}
\begin{tabular}{|c||ccc|c|}
\hline
& \multicolumn{3}{|c|}{${\rm{MSE}}(\{\tilde{\gamma}_j\}_{j\in\mathcal{J}})$} & \multirow{2}{*}{${\rm{MSE}}(\{\hat{\gamma}_j^{\rm{H}}\}_{j\in\mathcal{J}})$}\\
& Case 1: (\ref{Eq4.2.1}) & Case 1: (\ref{Eq4.2.2}) & Case 2: ${\bm{D}}={\bm{I}}_J$ &\\\hline\hline
$n=50$ & $1.06 \times 10^{-4}$ & $1.01 \times 10^{-4}$ & $27.47 \times 10^{-4}$ & $31.53 \times 10^{-4}$\\\hline
$n=200$ & $3.91 \times 10^{-5}$ & $3.73 \times 10^{-5}$ & $75.03 \times 10^{-5}$ & $77.82 \times 10^{-5}$\\\hline
\end{tabular}
\label{Table1}
\end{table}

Table \ref{Table1} shows results of predicting the EVIs $\{\gamma_j\}_{j\in\mathcal{J}}$ obtained using our method (\ref{Eq2.3.2.4}) and using the Hill estimator (\ref{Eq2.1.5}).
In this table, ${\rm{MSE}}(\{\tilde{\gamma}_j\}_{j\in\mathcal{J}})$ represents the mean square error (MSE) for our method (\ref{Eq2.3.2.4}), defined as
\begin{equation}
{\rm{MSE}}(\{\tilde{\gamma}_j\}_{j\in\mathcal{J}})\coloneqq\frac{1}{J\times M}\sum_{j=1}^J\sum_{m=1}^M\left(\tilde{\gamma}_j^{(m)}-\gamma_j\right)^2,\label{Eq4.2.3}
\end{equation}
where $\{\tilde{\gamma}_j^{(m)}\}_{j\in\mathcal{J}}$ are the predictors obtained from the $m$-th dataset.
Similarly, ${\rm{MSE}}(\{\hat{\gamma}_j^{\rm{H}}\}_{j\in\mathcal{J}})$ denotes the MSE for the Hill estimator (\ref{Eq2.1.5}).
As presented in Table \ref{Table1}, for all setups of ${\bm{D}}$, the MSE for our method (\ref{Eq2.3.2.4}) was improved compared to the MSE for the area-wise estimates obtained using the Hill estimator (\ref{Eq2.1.5}).
Particularly, appropriately adopting ${\bm{D}}$, such as (\ref{Eq4.2.1}) or (\ref{Eq4.2.2}), provided markedly greater improvement.
Consequently, in the next section, we deeply discuss some methods for constructing ${\bm{D}}$.
Even with the simplest setting ${\bm{D}}={\bm{I}}_J$, the proposed method performed slightly better than the area-wise estimation method did.
This performance suggests ${\bm{D}}={\bm{I}}_J$ as a candidate for the setup of ${\bm{D}}$.
These conclusions were confirmed for both cases $n=50$ and $n=200$.

\begin{figure}[t!]
\centering
\includegraphics[keepaspectratio, width=140mm]{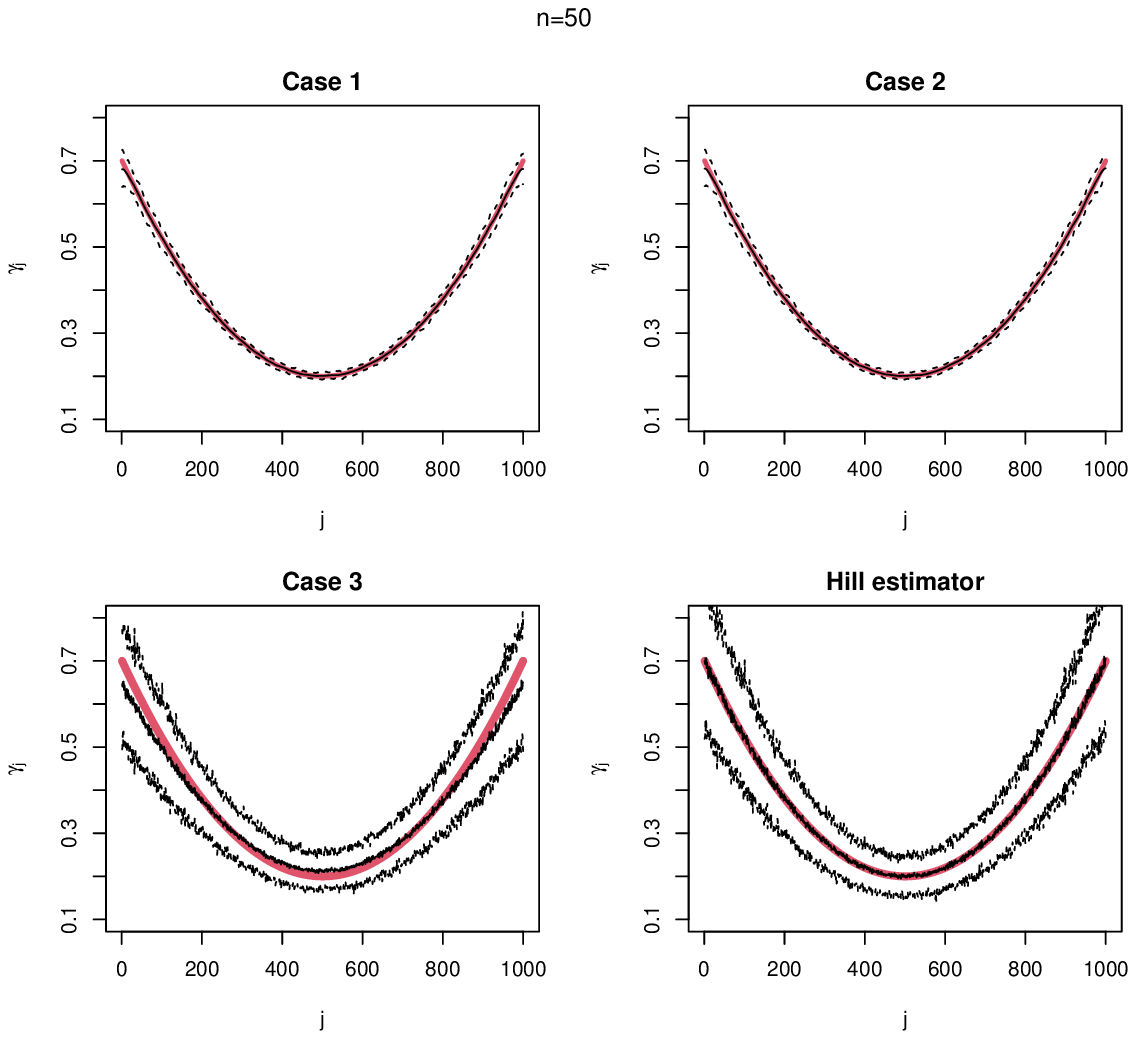}
\caption{Averages (solid line), 5th and 95th percentiles (dashed lines) of the predicted EVIs, and true EVIs (red line) for our methods with Case 1 (upper left panel), Case 2 (upper right panel), and Case 3 (lower left panel), and area-wise estimation (lower right panel), where $n=50$}\label{Figure4.2.1}
\end{figure}

\begin{figure}[t!]
\centering
\includegraphics[keepaspectratio, width=140mm]{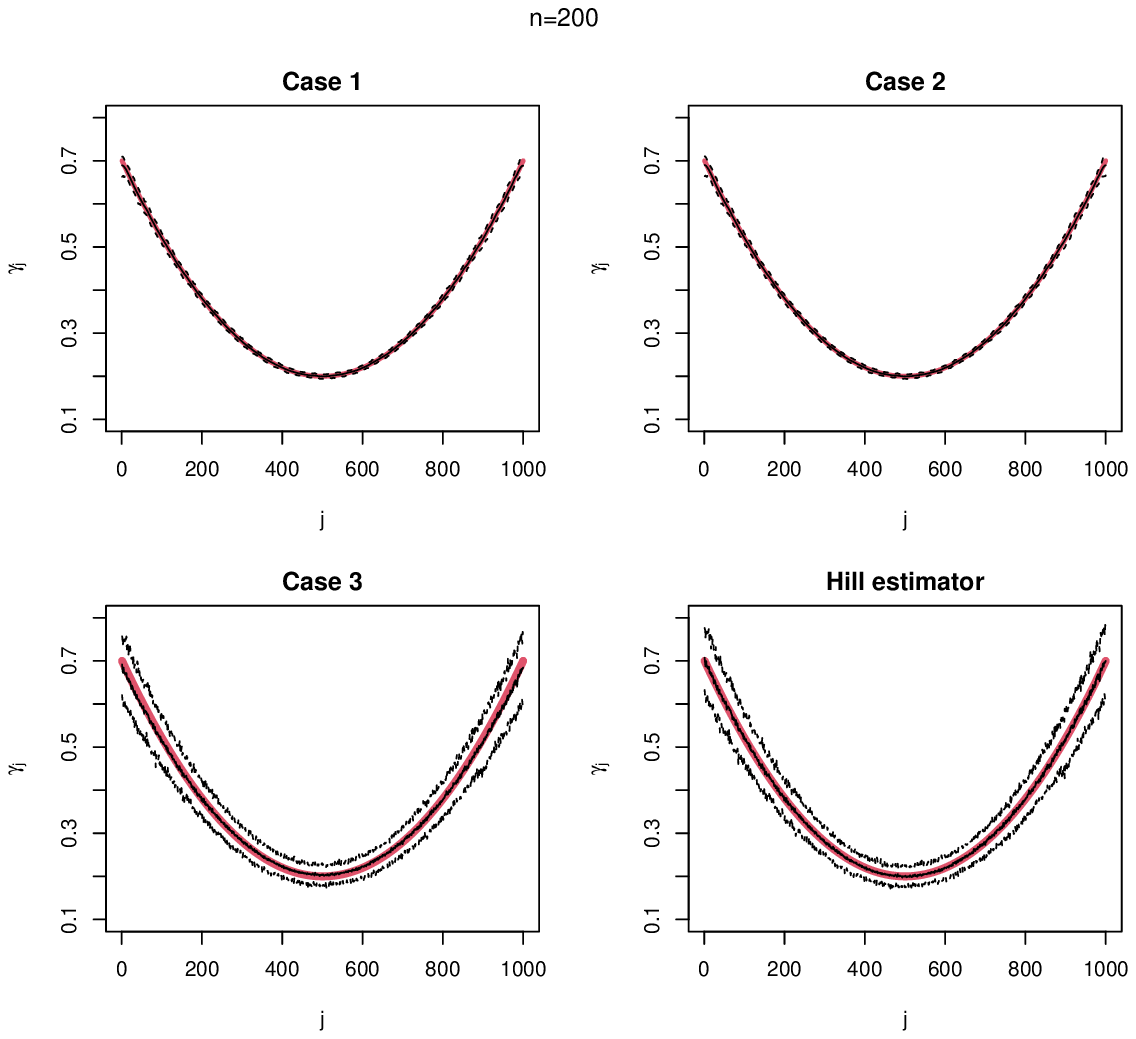}
\caption{Averages (solid line), 5th and 95th percentiles (dashed lines) of the predicted EVIs, and true EVIs (red line) for our methods with Case 1 (upper left panel), Case 2 (upper right panel), and Case 3 (lower left panel), and area-wise estimation (lower right panel), where $n=200$}\label{Figure4.2.2}
\end{figure}

Next, we examined the results presented in Table \ref{Table1} from the perspectives of bias and standard error.
For each method, we calculated the average and the 5th and 95th percentiles of the predicted EVIs in each area.
Figure \ref{Figure4.2.1} portrays the calculated averages (solid line), percentiles (dashed lines), and true EVIs (red line) for our methods with Case 1 (upper left panel), Case 2 (upper right panel), and Case 3 (lower left panel), and area-wise estimation (lower right panel), which correspond to the results obtained for $n=50$.
In the same manner, Figure \ref{Figure4.2.2} depicts results for $n=200$.
Figures \ref{Figure4.2.1} and \ref{Figure4.2.2} show that the predictors did not exhibit large biases, except for Case 3 with $n=50$.
For Case 3 with $n=50$, the predictors shifted toward $\hat{\mu}$ across areas and exhibited slight biases.
However, this phenomenon had the benefit of yielding more stable predictions, leading to the smaller MSE than area-wise estimation, as presented in Table \ref{Table1}.
Figures \ref{Figure4.2.1} and \ref{Figure4.2.2} presented significant differences in the standard errors of the predictors among methods.
Particularly, it is apparent that incorporating an appropriate ${\bm{D}}$, as in Cases 1 and 2, achieves highly stable predictions.

\section{Design of the Correlation Matrix of the Random Effects}\label{Section5}

As described in Sections \ref{Section2.3} and \ref{Section4.2}, the appropriate correlation matrix ${\bm{D}}$ of the random effects $\{V_j\}_{j\in\mathcal{J}}$ engenders significantly better performance of our method.
However, the random effects $\{V_j\}_{j\in\mathcal{J}}$ are latent variables and are not observed as data.
Therefore, we cannot evaluate ${\bm{D}}$ directly according to the definition (\ref{Eq2.2.3}).
Therefore, we must find ${\bm{D}}$ from other sources, while maintaining the fundamental interpretation of (\ref{Eq2.2.3}) that a large value of $D_{j_1j_2}=Cor[V_{j_1}, V_{j_2}], \ j_1\neq j_2\in\mathcal{J}$ denotes close values of the EVIs $\gamma_{j_1}=\exp(\mu+V_{j_1})$ and $\gamma_{j_2}=\exp(\mu+V_{j_2})$.
Below, we present three examples of methods for constructing ${\bm{D}}$.
\begin{itemize}
\item[(D1)]First, we consider the spatial data $\{Y_{ij}, i=1, 2, \ldots, n_j,\ j\in\mathcal{J}\}$.
In this case, the areas $\mathcal{J}$ refer to geographic sites, each of which has location information such as latitude and longitude.
For such spatial data, we can expect the EVIs $\{\gamma_j\}_{j\in\mathcal{J}}$ to be similar for areas that are close together.
Let ${\rm{lon}}_j$ and ${\rm{lat}}_j$ respectively denote the longitude and latitude of the area $j\in\mathcal{J}$.
Then, we can design the matrix ${\bm{D}}=[D_{j_1j_2}]_{j_1, j_2\in\mathcal{J}}$ as
\begin{equation*}
D_{j_1j_2}={\rm{ED}}_{j_1j_2}\coloneqq\exp\left[-\frac{{\rm{dis}}_{j_1j_2}}{c}\right]\in(0, 1],\quad j_1, j_2\in\mathcal{J},\label{Eq5.1.1}
\end{equation*}
where ${\rm{dis}}_{j_1j_2}$ is the Euclidean distance between areas $j_1\in\mathcal{J}$ and $j_2\in\mathcal{J}$, i.e., ${\rm{dis}}_{j_1j_2}\coloneqq\sqrt{\left({\rm{lon}}_{j_1}-{\rm{lon}}_{j_2}\right)^2+\left({\rm{lat}}_{j_1}-{\rm{lat}}_{j_2}\right)^2}$, and $c>0$ is a constant (Dyrrdal et al. 2015).
\item[(D2)] Regarding spatial Bayesian clustering for extreme value analysis, Rohrbeck and Tawn (2019) argued that ``{\it sites within the same cluster tend to exhibit a higher degree of dependence than sites in different clusters}'' and used the tail dependence (Reiss and Thomas 2007).
We also believe that areas with strong tail dependence have similar EVIs. We construct ${\bm{D}}$ based on the tail dependence as presented below.
We consider the data $\{Y_{ij},\ i=1, 2, \ldots, n,\ j\in\mathcal{J}\}$ as i.i.d. $J$-dimensional random vectors $\{(Y_{i1}, Y_{i2}, \ldots, Y_{iJ})^\top,\ i=1, 2, \ldots, n\}$ with the same joint distribution.
For each area $j\in\mathcal{J}$, the marginal distribution function of $\{Y_{ij}\}_{i=1}^{n_j}$ is denoted by $F_j(y)\coloneqq P(Y_{ij}\leq y)$.
Then, the tail dependence between two areas $j_1\in\mathcal{J}$ and $j_2\in\mathcal{J}$ is defined as ${\rm{TD}}_{j_1j_2}\coloneqq \lim_{p\uparrow 1}{\rm{TD}}_{j_1j_2}(p)$, where
\begin{equation*}
{\rm{TD}}_{j_1j_2}(p)\coloneqq P(Y_{ij_1}>F_{j_1}^{-1}(p)\mid Y_{ij_2}>F_{j_2}^{-1}(p)),\quad j_1, j_2\in\mathcal{J},\label{Eq5.1.2}
\end{equation*}
and $F_j^{-1}(\cdot)$ is the inverse function of $F_j(\cdot)$.
From Section 2.6 of Reiss and Thomas (2007), the above ${\rm{TD}}_{j_1j_2}$ is symmetric, i.e., ${\rm{TD}}_{j_1j_2}={\rm{TD}}_{j_2j_1}$.
Furthermore, we have $0\leq {\rm{TD}}_{j_1j_2}\leq 1$, indicating tail independence if ${\rm{TD}}_{j_1j_2}=0$ and total tail dependence if ${\rm{TD}}_{j_1j_2}=1$.
Roughly speaking, a strong tail dependence ${\rm{TD}}_{j_1j_2}$ suggests that extreme events in the two areas $j_1\in\mathcal{J}$ and $j_2\in\mathcal{J}$ are more likely to occur together.
The sample version of ${\rm{TD}}_{j_1j_2}$ is given as
\begin{equation*}
\widehat{\rm{TD}}_{j_1j_2}(p)\coloneqq\frac{1}{n(1-p)}\sum_{i=1}^nI_*(Y_{ij_1}>q_{j_1}(p), Y_{ij_2}>q_{j_2}(p))\label{Eq5.1.3}
\end{equation*}
for the fixed $p\approx 1$, where $q_j(p)$ is the $(100 \times p)\%$ empirical quantile of the data $\{Y_{ij}\}_{i=1}^{n_j}$, and $I_*(\cdot, \cdot)$ is the indicator function such that
\begin{equation*}
\begin{split}\label{Eq5.1.4}
&I_*(Y_{ij_1}>q_{j_1}(p), Y_{ij_2}>q_{j_2}(p))\\
&\quad\coloneqq
\begin{cases}
1, & Y_{ij_1}>q_{j_1}(p)\ {\rm{and}}\ Y_{ij_2}>q_{j_2}(p),\\
0, & {\rm{otherwise}}.
\end{cases}
\end{split}
\end{equation*}
Therefore, we use ${\bm{D}}=[\widehat{\rm{TD}}_{j_1j_2}(p)]_{j_1, j_2\in\mathcal{J}}$ as an alternative to $[{\rm{TD}}_{j_1j_2}]_{j_1, j_2\in\mathcal{J}}$, which does not include location information and which can therefore be applied to both spatial and non-spatial data.

\item[(D3)]
We assume that the given areas form some groups.
If areas $j_1\in\mathcal{J}$ and $j_2\in\mathcal{J}$ belong to different groups, then their correlation $D_{j_1, j_2}$ might be regarded as zero.
In this situation, the correlation matrix ${\bm{D}}$ is designed as a block matrix with zero submatrices.
Such a simpler construction of ${\bm{D}}$ can be expected to raise the efficiency of optimization process in our methods.
Even no clear group information for the given areas, we can impose a grouping structure on ${\bm{D}}$ such as (D1) and (D2) by setting small absolute values in  $[D_{j_1j_2}]_{j_1, j_2\in\mathcal{J}}$ to zero.
The simplest block structure of ${\bm{D}}$ is the identity matrix ${\bm{I}}_J$, which indicates that the data are unrelated across the areas.
However, ${\bm{D}}={\bm{I}}_J$ can still be used even when we have no information about ${\bm{D}}$.
As demonstrated in Section \ref{Section4.2}, this simple setup might also be more effective than area-wise estimation.
\end{itemize}

\section{Application to Extreme Precipitation in Japan}\label{Section6}

We demonstrated application of the proposed mixed effects model (\ref{Eq2.2.4}) by way of an example using a precipitation dataset from Japan.
The dataset is available on the Japan Meteorological Agency website (\url{https://www.data.jma.go.jp/gmd/risk/obsdl/index.php}), which includes records of daily precipitation (mm) kept for 1980--2022 at $J=1198$ weather stations (i.e., areas).
The period is equivalent to approximately $n=365 \times 43=15695$ days.
However, many stations lack records for consecutive dates according to their opening and closing dates.
We denote the dataset by $\{(Y_{i1}, Y_{i2}, \ldots, Y_{iJ})^\top,\ i=1, 2, \ldots, n\}$, where $Y_{ij}$ is the $i$-th daily precipitation in the area $j\in\mathcal{J}$, and the index $i$ represents the common date in all areas $\mathcal{J}$.
In Section \ref{Section6.1}, we first conducted an exploratory data analysis to determine the model setup for this dataset.
Section \ref{Section6.2} presents risk assessment for heavy rainfall in Japan using our mixed effects model.

\subsection{Preliminary analysis}\label{Section6.1}

Similarly to other studies of the spatial analysis of extreme precipitation, we did not consider temporal changes in extreme precipitation, thereby avoiding more complex models and uncertain results (Ragulina and Reitan 2017; Rohrbeck and Tawn 2021).

\begin{figure}[t!]
\centering
\includegraphics[keepaspectratio, width=120mm]{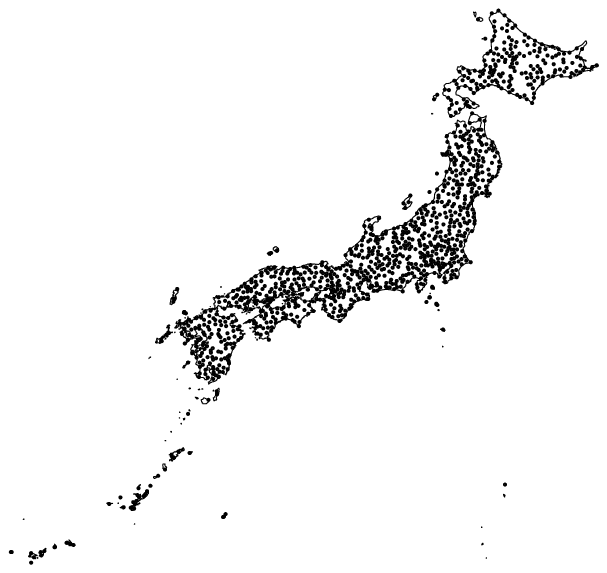}
\caption{Locations of $J=1138$ weather stations (areas) in Japan}
\label{Figure6.1.1}
\end{figure}

We first fitted the GPD to the data from each area $j\in\mathcal{J}$ as
\begin{equation}
\begin{split}\label{Eq6.1.1}
&P(Y_{ij}\leq y\mid Y_{ij}>\omega_j^{\rm{GPD}})\\
&=1-\left(1+\gamma_j^{\rm{GPD}}\frac{y-\omega_j^{\rm{GPD}}}{\psi_j^{\rm{GPD}}}\right)_+^{-1/\gamma_j^{\rm{GPD}}},\quad y>\omega_j^{\rm{GPD}},\ j\in\mathcal{J},
\end{split}
\end{equation}
where $\omega_j^{\rm{GPD}}$ is a given high threshold, $\gamma_j^{\rm{GPD}}$ and $\psi_j^{\rm{GPD}}$ are unknown parameters.
If $\gamma_j^{\rm{GPD}}\leq0$, then the assumption $\gamma_j>0$ in our model might be invalid (Chapter 1 of de Haan and Ferreira 2006).
Therefore, we first remove the areas with negative EVIs $\{\gamma_j^{\rm{GPD}}\}$ from the dataset.
To this end, we conducted the following hypothesis test for each area $j\in\mathcal{J}$ as
\begin{equation}
{\rm{H}}_{0j}: \gamma_j^{\rm{GPD}}>0\quad {\rm{vs.}}\quad {\rm{H}}_{1j}: \gamma_j^{\rm{GPD}}\leq 0,\label{Eq6.1.2}
\end{equation}
where ${\rm{H}}_{0j}$ represents the null hypothesis, and ${\rm{H}}_{1j}$ stands for the alternative hypothesis.
Let $\hat{\gamma}_j^{\rm{GPD}}$ be the maximum likelihood estimator of $\gamma_j^{\rm{GPD}}$ (Section 3.4 of de Haan and Ferreira 2006), where the threshold $\omega_j^{\rm{GPD}}$ in (\ref{Eq6.1.1}) was chosen using the GPD version of the discrepancy measure proposed by Wang and Tsai (2009).
Based on the test statistic given in Section 3.1 of Einmahl et al. (2019) and the asymptotic normality of $\hat{\gamma}_j^{\rm{GPD}}$, shown in Section 3.4 of de Haan and Ferreira (2006), we can reject the null hypothesis ${\rm{H}}_{0j}$ if $(k_j^{\rm{GPD}})^{1/2}\hat{\gamma}_j^{\rm{GPD}}\leq z_\alpha$, where $k_j^{\rm{GPD}}\coloneqq\sum_{i=1}^{n_j}I(Y_{ij}>\omega_j^{\rm{GPD}})$, and $z_\alpha$ is the $100 \times\alpha$-th percentile point of $N(0, 1)$, and $\alpha$ is a given significance level.
We applied the above hypothesis test (\ref{Eq6.1.2}) with $\alpha=0.05$ to each area $j\in\mathcal{J}$.
Based on the results, the null hypothesis ${\rm{H}}_{0j}$ was rejected for 60 areas.
Consequently, the subsequent analyses conducted using our proposed model omitted the 60 areas which were inferred to have negative EVI based on the above hypothesis tests, where $J$ was changed from $J=1198$ to $J=1138$.
Figure \ref{Figure6.1.1} shows the map with the locations of $J=1138$ areas.

\begin{figure}[t!]
\centering
\includegraphics[keepaspectratio, width=110mm]{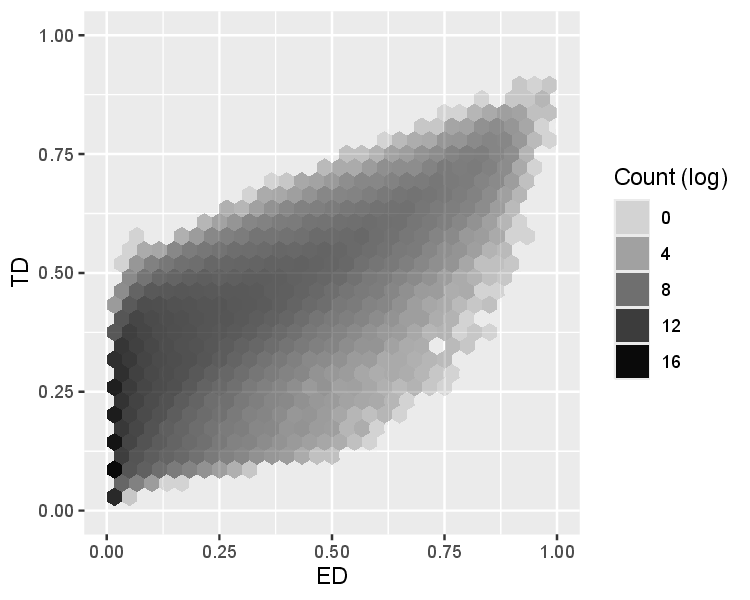}
\caption{Scatter plot of the Euclidean distance ${\rm{ED}}_{j_1j_2}$ versus tail dependence $\widehat{\rm{TD}}_{j_1j_2}(0.95)$ for all pairs of $j_1>j_2\in\mathcal{J}$.
The color of each hexagon represents the number of points within it}
\label{Figure6.1.2}
\end{figure}

To implement our model (\ref{Eq2.2.4}), we next constructed a correlation matrix ${\bm{D}}\in\mathbb{R}^{J\times J}$ of random effects.
Now, we can use both methods (D1) and (D2) described in Section \ref{Section5}.
Figure \ref{Figure6.1.2} shows a scatter plot of the Euclidean distance ${\rm{ED}}_{j_1j_2}$ versus tail dependence $\widehat{\rm{TD}}_{j_1j_2}(0.95)$ for all pairs of $j_1>j_2\in\mathcal{J}$.
As depicted in Figure \ref{Figure6.1.2}, the compositions of ${\bm{D}}=[{\rm{ED}}_{j_1j_2}]_{j_1, j_2\in\mathcal{J}}$ and ${\bm{D}}=[\widehat{\rm{TD}}_{j_1j_2}(0.95)]_{j_1, j_2\in\mathcal{J}}$ are similar, especially for elements close to 1.
Accordingly, we present the results only for ${\bm{D}}=[\widehat{\rm{TD}}_{j_1j_2}(0.95)]_{j_1, j_2\in\mathcal{J}}$, i.e., (D2).
Actually, the analysis results with (D1) did not differ greatly compared to those with (D2) for this application.

\subsection{The analysis by our mixed effects model}\label{Section6.2}

\begin{figure}[t!]
\centering
\includegraphics[keepaspectratio, width=100mm]{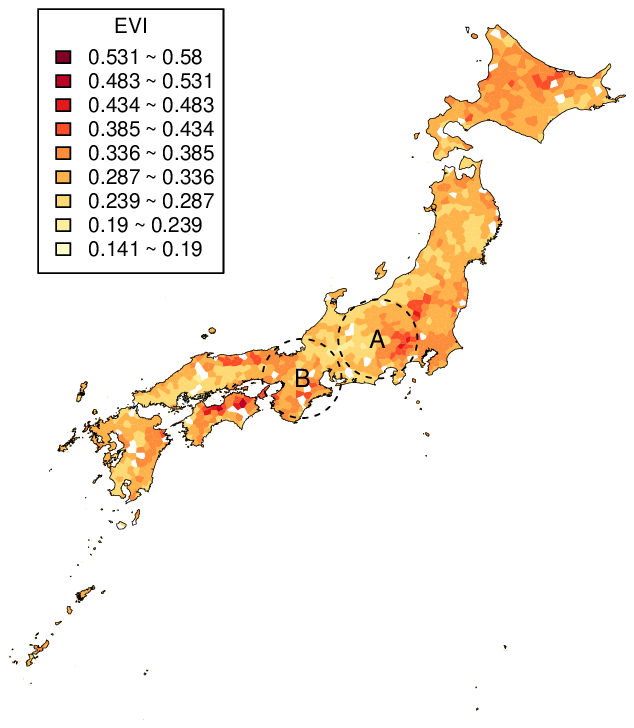}
\caption{Heatmap of the predicted EVIs $\{\tilde{\gamma}_j\}_{j\in\mathcal{J}}$ obtained using the proposed method}
\label{Figure6.2.1}
\end{figure}

\begin{figure}[t!]
\centering
\includegraphics[keepaspectratio, width=100mm]{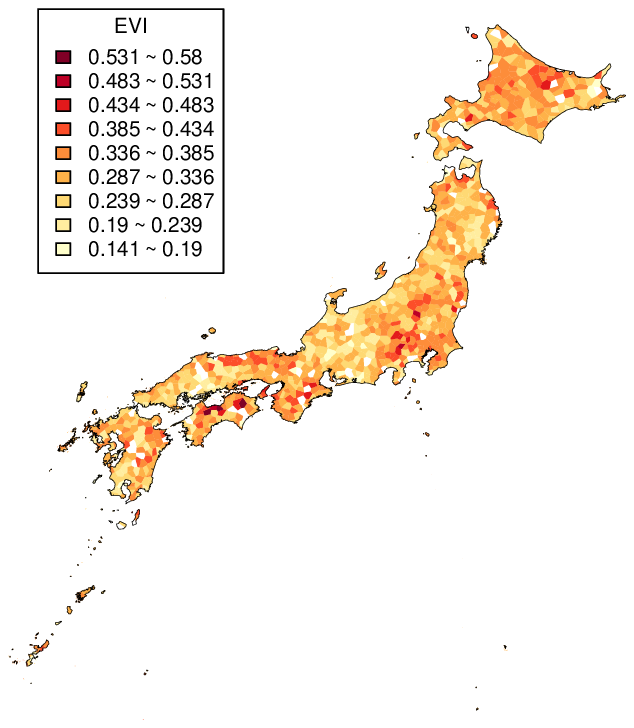}
\caption{Heatmap of the predicted EVIs $\{\hat{\gamma}_j^{\rm{H}}\}_{j\in\mathcal{J}}$ obtained using the Hill estimator}
\label{Figure6.2.5}
\end{figure}

\begin{figure}[t!]
\centering
\includegraphics[keepaspectratio, width=160mm]{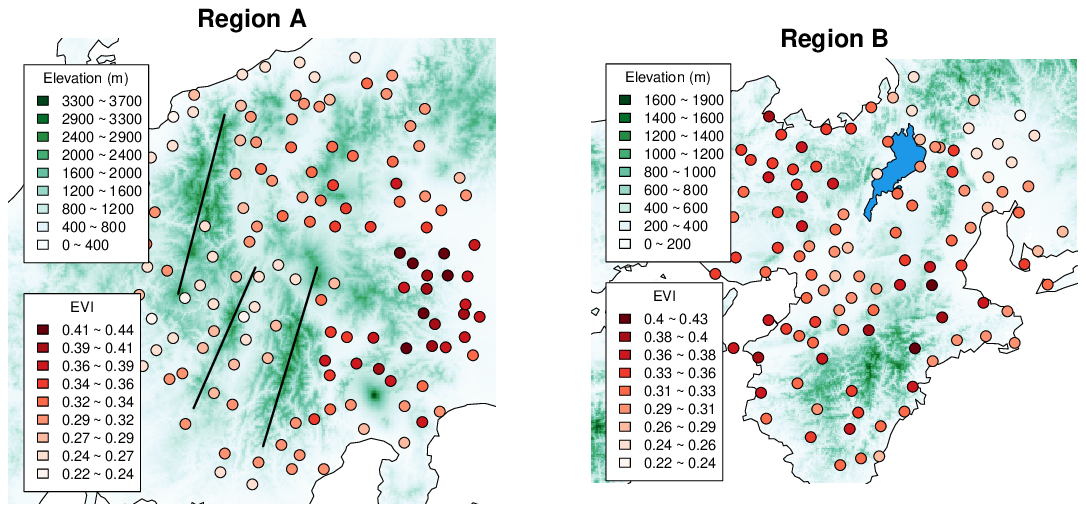}
\caption{Topographic maps of regions A and B in Figure \ref{Figure6.2.1}: Points show the locations of the areas and the predicted EVIs, the mountain ranges shown by three straight lines constitute the Japanese Alps; blue represents Lake Biwa}
\label{Figure6.2.2}
\end{figure}

To clarify the unknown structure of our model (\ref{Eq2.2.4}), we first obtained the estimates $\hat{\mu}$ and $\hat{\sigma}^2$, and predictor $\tilde{\bm{v}}=(\tilde{v}_1, \tilde{v}_2, \ldots, \tilde{v}_J)^\top\in\mathbb{R}^J$ for the thresholds $\{\omega_j\}_{j\in\mathcal{J}}$ chosen using a method similar to that described in Section \ref{Section6.1}.
The estimates were
\begin{equation*}
\hat{\mu}= -1.1642\ (\pm 0.1023)\quad {\rm{and}}\quad \hat{\sigma}^2= 0.0284\ (\pm0.0023),\label{Eq6.2.1}
\end{equation*}
where the value in parentheses represents the width of the 95\% confidence interval derived from the asymptotic normality in Theorem \ref{Theorem1}.
Figure \ref{Figure6.2.1} shows a heatmap of the predicted EVIs for all areas, i.e., $\tilde{\gamma}_j=\exp(\hat{\mu}+\tilde{v}_j),\ j\in\mathcal{J}$.
In this figure, the area boundaries were constructed using the package {\textsf{deldir}} (\url{https://CRAN.R-project.org/package=deldir}) within the {\textsf{R}}. The white area shows that the null hypothesis ${\rm{H}}_{0j}$ was rejected in Section \ref{Section6.1}.
Similarly, Figure \ref{Figure6.2.5} shows a heatmap of the predicted EVIs $\{\hat{\gamma}_j^{\rm{H}}\}$ obtained from area-wise estimation using the Hill estimator (Section \ref{Section2.1}).
Figures \ref{Figure6.2.1} and \ref{Figure6.2.5} suggest that our proposed method captured spatial variations in the EVIs more clearly than area-wise estimation using the Hill estimator.
This natural result of the proposed method comes from incorporating information about the relations between the areas through tail dependencies.
As shown in Figure \ref{Figure6.2.1}, some regions had local variations in the predicted EVIs.
Figure \ref{Figure6.2.2} highlights regions A and B in Figure \ref{Figure6.2.1} along with elevations, where the points denote the locations of the areas and their predicted EVIs.
Region A includes the mountain range called the Japanese Alps, whereas region B includes Lake Biwa, the largest lake in Japan.
From Figure \ref{Figure6.2.2}, it is apparent that in region A, the eastern side of the Japanese Alps had more large EVIs than the western side.
In region B, the EVIs were low around Lake Biwa and high in the southern mountains.

\begin{figure}[t!]
\centering
\includegraphics[keepaspectratio, width=100mm]{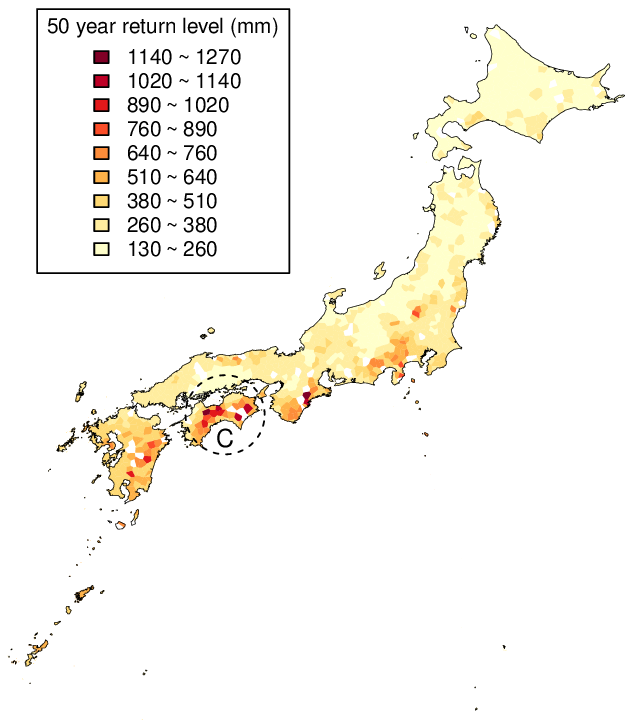}
\caption{Heatmap of the predicted 50-year return levels $\{\tilde{z}_j^{(50)}\}_{j\in\mathcal{J}}$}
\label{Figure6.2.3}

\includegraphics[keepaspectratio, width=120mm]{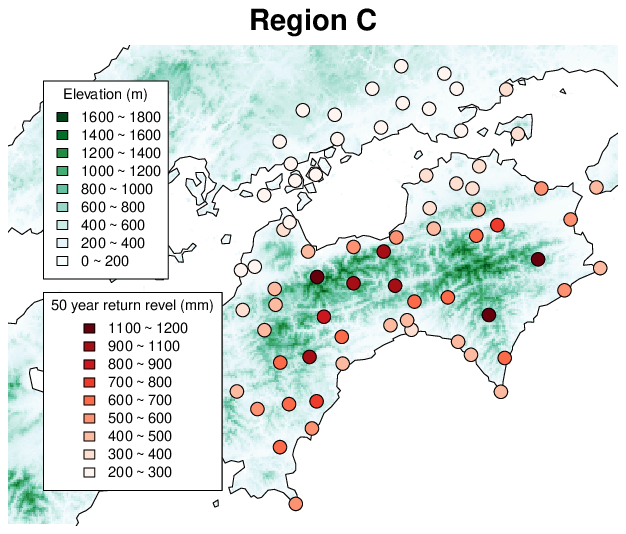}
\caption{Topographic map of region C in Figure \ref{Figure6.2.3}: Each point shows the locations of the areas as well as its predicted 50-year return levels}
\label{Figure6.2.4}
\end{figure}

Second, we predicted the high quantile of extreme rainfall for each area.
From (\ref{Eq2.2.4}) and (\ref{Eq2.3.1.3}), the frequency of extreme rainfall in each area can be evaluated as
\begin{equation}
\begin{split}\label{Eq6.2.2}
&P(Y_{ij}>y\mid V_j=\tilde{v}_j)\\
&\approx P(Y_{ij}>\omega_j\mid V_j=\tilde{v}_j)\left(\frac{y}{\omega_j}\right)^{-1/\tilde{\gamma}_j}\\
&\approx \frac{k_j}{n_j}\left(\frac{y}{\omega_j}\right)^{-1/\tilde{\gamma}_j},\quad j\in\mathcal{J}.
\end{split}
\end{equation}
For each area $j\in\mathcal{J}$, let $z_j^{(R)}$ be the $R$-year return level in the $j$-th area, which is defined as the high quantile which has probability $R^{-1}$ of being exceeded in a particular year (Cooley et al. 2007).
We can obtain $\{z_j^{(R)}\}_{j\in\mathcal{J}}$ by solving the equation $P(Y_{ij}>z_j^{(R)}\mid V_j=\tilde{v}_j)=(365R)^{-1}$ for each $j\in\mathcal{J}$.
Using the approximation (\ref{Eq6.2.2}), $\{z_j^{(R)}\}_{j\in\mathcal{J}}$ are predicted as
\begin{equation*}
\tilde{z}_j^{(R)}\coloneqq \omega_j\left(\frac{365Rk_j}{n_j}\right)^{\tilde{\gamma}_j},\quad j\in\mathcal{J}
\end{equation*}
(Chapter 4 of de Haan and Ferreira 2006).
Figure \ref{Figure6.2.3} depicts a heatmap of the predicted 50-year return levels $\{\tilde{z}_j^{(50)}\}_{j\in\mathcal{J}}$ for all areas.
From Figure \ref{Figure6.2.3}, it is apparent that the return levels tend to be higher on the Pacific side (i.e., the southwest of the map) because typhoons often affect these regions (Tro\v{s}elj and Lee 2021).
Figure \ref{Figure6.2.4} shows the topographic map of region C in Figure \ref{Figure6.2.3} with the predicted 50-year return levels, details of which are presented in the caption.
In region C, warm and humid air from the south becomes an upwelling because of the steep Shikoku Mountains, leading to more precipitation on the Pacific side, as described on the Japan Meteorological Agency website.
However, heavy rainfall is less likely to occur in the north over these mountains.
The large differences in return levels in Figure \ref{Figure6.2.4} reflected such weather conditions.
From Figure \ref{Figure6.2.4}, the highest 50-year return level among all areas was 1271 mm.
According to the dataset, the area with this highest return level experienced maximum total daily precipitation of 764 mm during 1980--2022.
Therefore, in such an area, we should be alert to the occurrence of unprecedented and heavier rainfalls.

\section{Discussion}\label{Section7}

For extreme value analysis of data from multiple areas, we studied the new POT method using the mixed effects model.
Because the area-wise effective sample sizes are often small in extreme value analysis, classical estimation methods such as the Hill estimator tend to yield uncertain results.
Consequently, this study used information about the relations between the areas.
Such a model was achieved using the mixed effects model with correlated random effects.
According to the asymptotic theory in Section \ref{Section3}, the mixed effects model supports our motivation to use richer area information.
In fact, the performance of the estimators for our model improves as the number of areas, $J$, increases (R2 in Section \ref{Section3}).
Therefore, as explained in the application in Section \ref{Section6}, our model is useful for analyzing large data with many areas.
Furthermore, our numerical experiment explained in Section \ref{Section4.2} demonstrated that appropriately correlated random effects of the mixed effects model improve area-wise estimates in extreme value analysis considerably.
In the application in Section \ref{Section6}, our method provided a more natural analytical result than the area-wise estimation method (Figures \ref{Figure6.2.1} and \ref{Figure6.2.5}).
Some examples for constructing the correlation matrix of the random effects are discussed in Section \ref{Section5}.

For this study, the POT method using the mixed effects model was provided for the Pareto-type distribution.
More generally, it might also be developed for the GPD explained in Section \ref{Section2.1}.
However, such an extension involves the following challenges.
First, because of the POT method specifications, the scale parameter in the GPD is typically related to the EVI and threshold in the GPD (Theorem 1.2.5 of de Haan and Ferreira 2006).
For example, if the EVIs are assumed to be a mixed effects model, then the scale parameters also affect the random effects.
Therefore, the mixed effects modeling for the GPD might be more complex than the Pareto-type model.
Second, in the mixed-effects model with the GPD, the EVI can be any real number, although it must be greater than $-1/2$ to ensure stability. 
On the other hand, the behavior of the scale parameter depends on the sign of the EVI (Theorem 1.2.5 of de Haan and Ferreira 2006).
Therefore, when the likelihood function defined as (\ref{Eq2.3.1.4}) replaced the Pareto-type density with the density of the GPD, the integral of the random effects becomes discontinuous at the points where the EVIs equal zero.
This discontinuity might complicate the optimization process for maximizing the log-likelihood.
These difficulties might be avoided if the EVIs of all areas have a common sign (positive or negative).
Our results presented herein correspond to the case in which all EVIs are positive. 
If all areas were to have negative EVIs, then our method would be able to be developed for the GPD case.
Then, for example, the EVIs are modeled as $\gamma_j=-2^{-1}(1+\exp[\mu+V_j])^{-1},\ j\in\mathcal{J}$ such that $-1/2<\gamma_j<0$.
This is left as an important study for future work.

\section*{Appendix}\label{Appendix}

This appendix presents key results necessary for proving Theorem \ref{Theorem1}.
Let
\begin{equation*}
l({\bm{v}}; \mu)\coloneqq(Jk)^{-1}\sum_{j\in\mathcal{J}}\sum_{i=1, 2, \ldots, n_j: Y_{ij}>\omega_j}\left[\mu+v_j+\left\{\frac{1}{\exp(\mu+v_j)}+1\right\}\log\frac{Y_{ij}}{\omega_j}\right].
\end{equation*}
We denote $\dot{\bm{v}}=(\dot{v}_1, \dot{v}_2, \ldots, \dot{v}_J)^\top\coloneqq {\rm{argmin}}_{{\bm{v}}\in\mathbb{R}^J}l({\bm{v}}; \mu)$.
Similarly to (35) of Momoki and Yoshida (2025), we have that under conditioning on ${\bm{V}}={\bm{v}}$,
\begin{equation}
\dot{v}_j=v_j+J\frac{k}{k_j}\frac{\partial}{\partial v_j}l({\bm{v}}; \mu)\left(1+o_P(1)\right), \quad j\in\mathcal{J}.\label{A3}
\end{equation}
Likelihood function $L(\mu, \sigma^2)$ satisfies
\begin{equation}
\begin{split}
&\frac{\partial}{\partial\mu}\log L(\mu, \sigma^2)\\
&=\frac{\int_{\mathbb{R}^J}\frac{\partial}{\partial \mu}\phi_J({\bm{v}}; \mu{\bm{1}}_J, \sigma^2{\bm{D}})\exp\left[-\left(Jk\right)l({\bm{v}}-\mu{\bm{1}}_J; \mu)\right]d{\bm{v}}}{\int_{\mathbb{R}^J}\phi_J({\bm{v}}; \mu{\bm{1}}_J, \sigma^2{\bm{D}})\exp\left[-\left(Jk\right)l({\bm{v}}-\mu{\bm{1}}_J; \mu)\right]d{\bm{v}}}\label{A7}
\end{split}
\end{equation}
and
\begin{equation}
\begin{split}
&\frac{\partial}{\partial\sigma^2}\log L(\mu, \sigma^2)\\
&=\frac{\int_{\mathbb{R}^J}\frac{\partial}{\partial\left(\sigma^2\right)}\phi_J({\bm{v}}; \mu{\bm{1}}_J, \sigma^2{\bm{D}})\exp\left[-\left(Jk\right)l({\bm{v}}-\mu{\bm{1}}_J; \mu)\right]d{\bm{v}}}{\int_{\mathbb{R}^J}\phi_J({\bm{v}}; \mu{\bm{1}}_J, \sigma^2{\bm{D}})\exp\left[-\left(Jk\right)l({\bm{v}}-\mu{\bm{1}}_J; \mu)\right]d{\bm{v}}},\label{A8}
\end{split}
\end{equation}
where we have
\begin{equation}
\frac{\partial}{\partial \mu}\phi_J({\bm{v}}; \mu{\bm{1}}_J, \sigma^2{\bm{D}})={\bm{1}}_J^\top\left(\sigma^2{\bm{D}}\right)^{-1}\left({\bm{v}}-\mu{\bm{1}}_J\right)\phi_J({\bm{v}}; \mu{\bm{1}}_J, \sigma^2{\bm{D}})\label{A1}
\end{equation}
and
\begin{equation}
\begin{split}
&\frac{\partial}{\partial\left(\sigma^2\right)}\phi_J({\bm{v}}; \mu{\bm{1}}_J, \sigma^2{\bm{D}})\\
&=\left(\sigma^2\right)^{-1}\left[\frac{\left({\bm{v}}-\mu{\bm{1}}_J\right)^\top\left(\sigma^2{\bm{D}}\right)^{-1}\left({\bm{v}}-\mu{\bm{1}}_J\right)-1}{2}\right]\phi_J({\bm{v}}; \mu{\bm{1}}_J, \sigma^2{\bm{D}}).\label{A2}
\end{split}
\end{equation}
We use (\ref{A3}) as
\begin{equation}
\begin{split}
&\left({\bm{1}}_J^\top{\bm{D}}^{-1}{\bm{1}}_J\right)^{-1/2}{\bm{1}}_J^\top\left(\sigma^2{\bm{D}}\right)^{-1}\left\{\left(\dot{\bm{v}}+\mu{\bm{1}}\right)-\mu{\bm{1}}_J\right\}\\
&=\left({\bm{1}}_J^\top{\bm{D}}^{-1}{\bm{1}}_J\right)^{-1/2}{\bm{1}}_J^\top\left(\sigma^2{\bm{D}}\right)^{-1}{\bm{V}}\\
&\quad+\left({\bm{1}}_J^\top{\bm{D}}^{-1}{\bm{1}}_J\right)^{-1/2}{\bm{1}}_J^\top\left(\sigma^2{\bm{D}}\right)^{-1}\left[J\frac{k}{k_j}\frac{\partial}{\partial v_j}l({\bm{V}}; \mu)\right]_{j\in\mathcal{J}}\left(1+o_P(1)\right).\label{A4}
\end{split}
\end{equation}
From (\ref{Eq2.2.3}), the first term on the right-hand side of (\ref{A4}) converges to $N(0, (\sigma^2)^{-1})$ in distribution as $J\to\infty$.
For the second term on the right-hand side of (\ref{A4}), we have
\begin{equation}
\begin{split}
&\left({\bm{1}}_J^\top{\bm{D}}^{-1}{\bm{1}}_J\right)^{-1/2}{\bm{1}}_J^\top\left(\sigma^2{\bm{D}}\right)^{-1}\left[J\frac{k}{k_j}\frac{\partial}{\partial v_j}l({\bm{V}}; \mu)\right]_{j\in\mathcal{J}}\\
&\quad-\left({\bm{1}}_J^\top{\bm{D}}^{-1}{\bm{1}}_J\right)^{-1/2}{\bm{1}}_J^\top\left(\sigma^2{\bm{D}}\right)^{-1}\left[E\left[\varepsilon_j(V_j)\right]\right]_{j\in\mathcal{J}}\\
&=k^{-1/2}\left({\bm{1}}_J^\top{\bm{D}}^{-1}{\bm{1}}_J\right)^{-1/2}{\bm{1}}_J^\top\left(\sigma^2{\bm{D}}\right)^{-1}\\
&\quad \times\left[\left(\frac{k_j}{k}\right)^{-1/2}\left\{Jkk_j^{-1/2}\frac{\partial}{\partial v_j}l({\bm{V}}; \mu)-k_j^{1/2}\varepsilon_j(V_j)\right\}\right]_{j\in\mathcal{J}}\\
&\quad\quad +\left({\bm{1}}_J^\top{\bm{D}}^{-1}{\bm{1}}_J\right)^{-1/2}{\bm{1}}_J^\top\left(\sigma^2{\bm{D}}\right)^{-1}\left[\varepsilon_j(V_j)-E\left[\varepsilon_j(V_j)\right]\right]_{j\in\mathcal{J}},\label{A5}
\end{split}
\end{equation}
where the right-hand side of (\ref{A5}) converges to 0 in probability as $n_j\to\infty,\ j\in\mathcal{J}$ and $J\to\infty$ (Lemma 2 of Momoki and Yoshida 2025).
Consequently, we obtain
\begin{equation}
\begin{split}
&\left({\bm{1}}_J^\top{\bm{D}}^{-1}{\bm{1}}_J\right)^{-1/2}{\bm{1}}_J^\top\left(\sigma^2{\bm{D}}\right)^{-1}\left\{\left(\dot{\bm{v}}+\mu{\bm{1}}\right)-\mu{\bm{1}}_J\right\}\\
&\quad-\left({\bm{1}}_J^\top{\bm{D}}^{-1}{\bm{1}}_J\right)^{-1/2}{\bm{1}}_J^\top\left(\sigma^2{\bm{D}}\right)^{-1}\left[E\left[\varepsilon_j(V_j)\right]\right]_{j\in\mathcal{J}}\\
&\quad\quad\xrightarrow{D}N\left(0, \left(\sigma^2\right)^{-1}\right)\label{A9}
\end{split}
\end{equation}
as $n_j\to\infty,\ j\in\mathcal{J}$ and $J\to\infty$.
Furthermore, from (\ref{A2}), we can write as
\begin{equation}
\begin{split}
&J^{-1/2}\left(\sigma^2\right)^{-1}\left[\frac{\left\{\left(\dot{\bm{v}}+\mu{\bm{1}}\right)-\mu{\bm{1}}_J\right\}^\top\left(\sigma^2{\bm{D}}\right)^{-1}\left\{\left(\dot{\bm{v}}+\mu{\bm{1}}\right)-\mu{\bm{1}}_J\right\}-1}{2}\right]\\
&=2^{-1/2}\left(\sigma^2\right)^{-1}\left[\frac{{\bm{V}}^\top\left(\sigma^2{\bm{D}}\right)^{-1}{\bm{V}}-1}{\left(2J\right)^{1/2}}\right]\\
&\quad+J^{-1/2}\left(\sigma^2\right)^{-2}{\bm{V}}^\top{\bm{D}}^{-1}\left[J\frac{k}{k_j}\frac{\partial}{\partial v_j}l({\bm{V}}; \mu)\right]_{j\in\mathcal{J}}\left(1+o_P(1)\right).\label{A6}
\end{split}
\end{equation}
Because the random variable ${\bm{V}}^\top(\sigma^2{\bm{D}})^{-1}{\bm{V}}$ has the chi-squared distribution $\chi_J$ with $J$ degrees of freedom, the first term on the right-hand side of (\ref{A6}) converges to $N(0, 2^{-1}(\sigma^2)^{-2})$ in the distribution as $J\to\infty$.
Similar to the second term on the right-hand side of (\ref{A4}), the second term on the right-hand side of (\ref{A6}) is asymptotically equivalent to $J^{-1/2}(\sigma^2)^{-2}E[{\bm{V}}^\top{\bm{D}}^{-1}[\varepsilon_j(V_j)]_{j\in\mathcal{J}}]$ as $n_j\to\infty,\ j\in\mathcal{J}$ and $J\to\infty$.
Therefore, we obtain
\begin{equation}
\begin{split}
&J^{-1/2}\left(\sigma^2\right)^{-1}\left[\frac{\left\{\left(\dot{\bm{v}}+\mu{\bm{1}}\right)-\mu{\bm{1}}_J\right\}^\top\left(\sigma^2{\bm{D}}\right)^{-1}\left\{\left(\dot{\bm{v}}+\mu{\bm{1}}\right)-\mu{\bm{1}}_J\right\}-1}{2}\right]\\
&\quad-J^{-1/2}\left(\sigma^2\right)^{-2}E\left[{\bm{V}}^\top{\bm{D}}^{-1}\left[\varepsilon_j(V_j)\right]_{j\in\mathcal{J}}\right]\\
&\quad\quad\xrightarrow{D}N\left(0, 2^{-1}\left(\sigma^2\right)^{-2}\right)\label{A10}
\end{split}
\end{equation}
as $n_j\to\infty,\ j\in\mathcal{J}$ and $J\to\infty$.
We apply the Laplace approximation in Appendix A of Miyata (2004) to (\ref{A7}) and (\ref{A8}).
Using (\ref{A9}) and (\ref{A10}), we then obtain
\begin{equation}
\begin{split}
&\left({\bm{1}}_J^\top{\bm{D}}^{-1}{\bm{1}}_J\right)^{-1/2}\frac{\partial}{\partial\mu}\log L(\mu, \sigma^2)\\
&\quad-\left({\bm{1}}_J^\top{\bm{D}}^{-1}{\bm{1}}_J\right)^{-1/2}{\bm{1}}_J^\top\left(\sigma^2{\bm{D}}\right)^{-1}\left[E\left[\varepsilon_j(V_j)\right]\right]_{j\in\mathcal{J}}\\
&\quad\quad\xrightarrow{D}N\left(0, \left(\sigma^2\right)^{-1}\right)
\end{split}
\end{equation}
and
\begin{equation}
\begin{split}
&J^{-1/2}\frac{\partial}{\partial\left(\sigma^2\right)}\log L(\mu, \sigma^2)-J^{-1/2}\left(\sigma^2\right)^{-2}E\left[{\bm{V}}^\top{\bm{D}}^{-1}\left[\varepsilon_j(V_j)\right]_{j\in\mathcal{J}}\right]\\
&\quad\xrightarrow{D}N\left(0, 2^{-1}\left(\sigma^2\right)^{-2}\right)
\end{split}
\end{equation}
as $n_j\to\infty,\ j\in\mathcal{J}$ and $J\to\infty$.
The remainder of the proof of Theorem \ref{Theorem1} resembles the proof of Theorem 1 presented by Momoki and Yoshida (2025).

\section*{Declarations}

\noindent
{\bf{Competing interests}}

The authors have no relevant financial or non-financial interests to disclose.

\noindent
{\bf{Funding}}

This work was supported by JSPS KAKENHI (Grant Nos. 22K11935 and 23K28043).

\noindent
{\bf{Authors' contributions}}

All authors contributed to the study conception and design. Methodology, investigation, and data analysis were mainly performed by Koki Momoki. The manuscript was mainly written by Koki Momoki. All authors reviewed the manuscript and approved the final version of the manuscript.

\noindent
{\bf{Acknowledgments}}

We would like to thank FASTEK (\url{https://www.fastekjapan.com/}) for English language editing. This research was financially supported by JSPS KAKENHI (Grant Nos. 22K11935 and 23K28043).

\noindent
{\bf{Data availability}}

The dataset analyzed during this study was processed from those obtained from the Japan Meteorological Agency website (\url{https://www.data.jma.go.jp/gmd/risk/obsdl/index.php}), which is available from the corresponding author on reasonable request. The code of \textsf{R} (\url{https://www.r-project.org/}) for the simulation studies is also available from the corresponding author on reasonable request.

\subsection*{Compliance with Ethical Standards}
{\bf{Disclosure of potential conflicts of interest}} Not applicable.\\
\noindent
{\bf{Research involving Human Participants and/or Animals}} Not applicable.\\
\noindent
{\bf{Informed consent}} Not applicable.

\end{document}